\documentclass[a4paper]{jpconf}
\usepackage{graphicx,iopams}
\begin{document}
\title{The stellar cusp around the Milky Way's central black hole}

\author{R. Sch{\"o}del$^{1}$, E. Gallego-Cano$^{1}$ and P. Amaro-Seoane$^{2}$}

\address{$^{1}$Instituto de Astrof{\'i}sica de Andaluc{\'i}a (CSIC),
  Glorieta de la Astronom{\'i}a s/n, 18008 Granada, Spain\\
  $^{2}$Institut de Ci{\`e}ncies de l'Espai (CSIC-IEEC) at Campus UAB, Carrer de Can Magrans s/n 08193 Barcelona, Spain\\
Institute of Applied Mathematics, Academy of Mathematics and Systems Science, CAS, Beijing 100190, China\\
Kavli Institute for Astronomy and Astrophysics, Beijing 100871, China\\
Zentrum f{\"u}r Astronomie und Astrophysik, TU Berlin, Hardenbergstra{\ss}e 36, 10623 Berlin, Germany
}

\ead{rainer@iaa.es}

\begin{abstract}
  The existence of stellar cusps in dense clusters around massive
  black holes is a fundamental, decades-old prediction of theoretical
  stellar dynamics. Yet, observational evidence has been difficult to
  obtain. With a new, improved analysis of high-angular resolution
  images of the central parsecs of the Galactic Center, we are finally
  able to provide the first solid evidence for the existence of a
  stellar cusp around the Milky Way's massive black hole. The
  existence of stellar cusps has a significant impact on predicted
  event rates of phenomena like tidal disruptions of stars and extreme
  mass ratio inspirals.
\end{abstract}

\section{Introduction}

It is a standard paradigm of modern astrophysics that the majority of
galaxies -- possibly with the exception of very low mass galaxies and/or dwarf
irregulars -- contains massive black holes (MBHs) at their centres (see,
e.g., \cite{Gultekin:2009fk} and references therein). Over the past two
decades, mainly thanks to studies with the Hubble Space Telescope and
high-angular resolution observations with Adaptive Optics at major
ground-based telescopes, we have also learned that the majority of
galaxies contains so-called {\it nuclear star clusters} (NSCs) at their
photometric and dynamical centres. These clusters are the densest and
most massive clusters that can be found in the present-day
Universe. They have masses between a few $10^{5}$ to
$10^{8}$\,M$_{\odot}$ and half light radii of a few to a few tens of
parsecs. NSCs do not consist of single-age stellar populations, but
rather show signs of repeated star formation along their host
galaxies' lifetime, i.e.\ they are in their properties significantly
different from globular clusters. Most intriguingly, NSCs have been
found to coexist with massive black holes (for the properties of NSCs,
see \cite{Boker:2010ys,Georgiev:2014ve,Georgiev:2016nr} and references therein). 

The evolution of a star cluster containing an MBH is a decades-old
problem of stellar dynamics that has been studied by a large number of
authors with a wide range of techniques (analytical, Fokker-Planck,
Nbody). Peebles (1972) \cite{Peebles:1972fk} proved that, while the
statistical thermal equilibrium must be violated close to a MBH in a
galactic nucleus due to tidal disruptions, stellar collisions and
gravitational captures, there exists a steady state with net inward
flux of stars.  This is a quasi-steady state solution,
where the stellar density takes a power-law form
$\rho \propto r^{-\gamma}$. This finding was corroborated for a
single-mass population by Bahcall and Wolf \cite{Bahcall:1976vn} and
subsequently for multi-mass stellar populations with realistic number
fractions (e.g.,
\cite{Bahcall:1977ys,Lightman:1977ly,Murphy:1991zr,Amaro-Seoane:2004kx,Alexander:2005fk,Merritt:2006ys,Alexander:2009gd,Amaro-Seoane:2011qv}
and references therein). This so-called stellar density {\it cusp}
will be fully developed after a so-called {\it relaxation time}, the
time necessary for the randomisation of the cluster phase space via
close encounters between stars (so-called two-body relaxation), which
is typically below a Hubble time for a NSC in a Milky Way-like galaxy
For a cluster composed only of stars of a single mass the predicted
value is $\gamma=1.75$. For realistic, multi-mass clusters, the lower
mass stars have $\gamma=1.5$, while the heavier stars (and in
particular stellar mass black holes) will follow a steeper
distribution ($\gamma_{BH}\approx2$; see
\cite{Alexander:2009gd,Amaro-Seoane:2011qv}). The cusp will be well
developed inside the radius of influence of the MBH, which is roughly
the radius of a sphere that contains once to twice the black hole mass
in form of stars (or stellar remnants) \cite{Alexander:2005fk}.

Although stellar cusps around MBHs are a robust theoretical
prediction, they have been rather elusive observationally. That is
mainly because of the great distances of MBHs: Their radii of
influence are often resolved only by a few pixels in imaging or
spectroscopy, which means that a handful of rare bright stars can bias
the measured cluster structure significantly, as can be seen very well
in the case of our own Milky Way (see discussion in
\cite{Schodel:2007tw}). When comparing theoretical predictions with
observations in nature, it is also important to be aware that
theoretical models are often highly simplified, e.g.\ they typically
are setup with clusters of a single age stellar population, and do not
suffer from observational effects such as strong and spatially
variable interstellar extinction.

\begin{figure}[htb]
\label{Fig:overview}
\includegraphics[width=\textwidth]{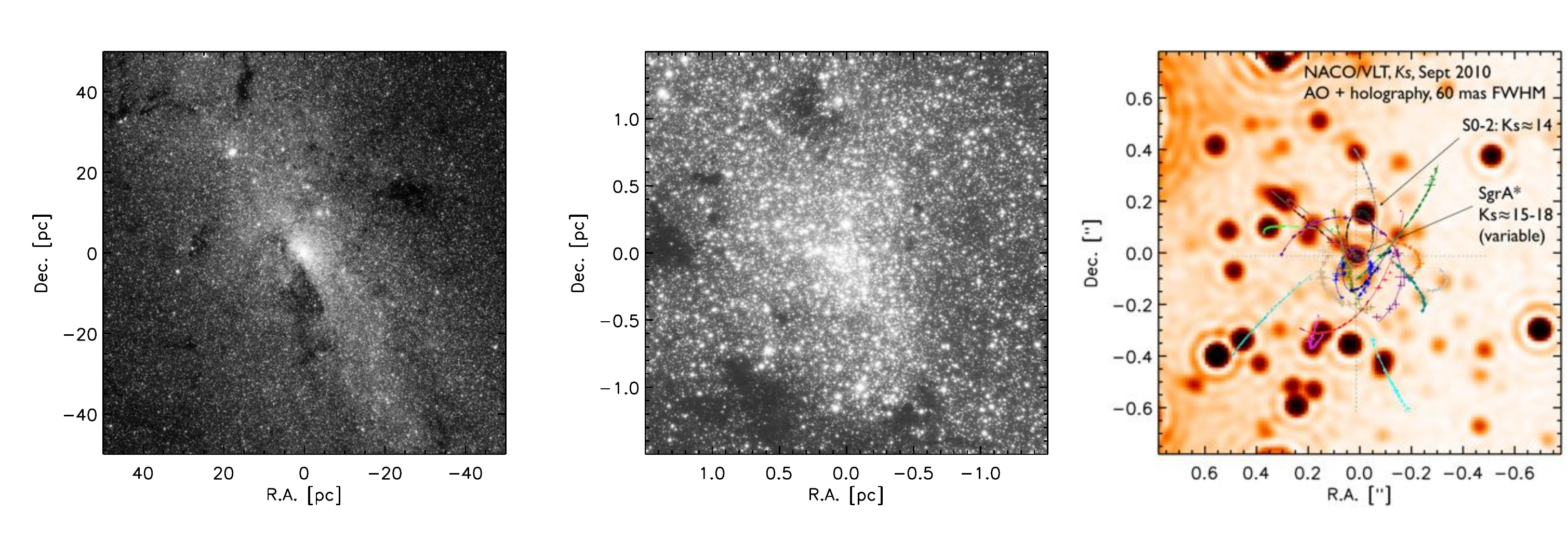}
\caption{A progressive zoom into the Galactic Center. Left: Spitzer
  $4.5\,\mu$m image \cite{Stolovy:2006fk} of the central
  $100\,\mathrm{pc}\times100\,\mathrm{pc}$ of the Galaxy. Coordinates
  are given as offsets from the MBH Sgr\,A*. The NSC is the
  bright, compact structure in the centre of the image. Middle:
  NACO/VLT $2.2\,\mu$m image of the central
  $3\,\mathrm{pc}\times3\,\mathrm{pc}$ around Sgr\,A*. Right: NACO/VLT
  $2.2\,\mu$m image of the central $1.6"\times1.6"$
  ($0.06\,\mathrm{pc}\times0.06\,\mathrm{pc}$) around Sgr\,A*. Stellar
  orbits determined by \cite{Gillessen:2016uq} are over-plotted onto
  the image. Sgr\,A* is visible as a near-infrared source at the image
centre.}
\end{figure}

The best case for testing the stellar cusp hypothesis is the centre of
the Milky Way. In the Galactic Centre (GC), an MBH of about
$4\times10^{6}$\,M$_{\odot}$, called Sagittarius\,A* (Sgr\,A*), is
surrounded by an NSC of about $2.5\times10^{7}$\,M$_{\odot}$, at a
distance of $\sim$8\,kpc (\cite{Genzel:2010fk,Boehle:2016zr,
  Gillessen:2016uq,Schodel:2014fk} and references therein). With high
angular resolution observations we can therefore probe the cluster
structure and dynamics on scales of milli-parsecs, which means that
the GC is a prime laboratory for studying the interaction of a stellar
cluster with an MBH \cite{Schodel:2014bn}. Figure\,\ref{Fig:overview}
provides an overview of the GC and a progressive zoom toward the MBH
Sagittarius\,A* (Sgr\,A*).

It is important to keep in mind the limitations on observational
studies of the GC: (1) Due to the extremely high interstellar
extinction, stars can only be detected with reasonable sensitivity in
the near-infrared. In this region, however, intrinsic stellar colours
are small and in combination with the strong differential reddening
caused by the small-scale variability of extinction it is therefore
very hard to classify the stars. For example, it is not trivial to
distinguish between an old, cold giant star and a young, hot massive
star by means of imaging and stellar colours. (2) Due to the extreme
density of the NSC, one needs to work with high angular resolution,
typically at the diffraction limit of ground-based 8-10m-class
telescopes, which can only be achieved with special techniques
(Adaptive Optics and speckle imaging). Even then, most faint stars
(main sequence stars of less than two solar masses) cannot be detected
because of the source crowding. Hence, we can only observe the tip of
the iceberg, on the order a few to 10\% of all the stars suspected to
exist at the GC.

Because of the observational limitations, evidence on the existence of
a stellar cusp at the GC was elusive. Although the detection of a cusp
was claimed by first high angular resolution observations
\cite{Schodel:2007tw,Genzel:2003it}, the actual situation turned out
to be more complex. Only stars older than the cluster relaxation time
can serve as suitable tracers of a cusp, but there are many massive
young stars present in the central parsec around Sgr\,A*. These are
too young to be dynamically relaxed. Once the pollution from young
stars was taken into account, it appeared that the stellar surface
number density within a few 0.1\,pc of the MBH at the GC was close to
flat, or even decreasing
\cite{Buchholz:2009fk,Do:2009tg,Bartko:2010fk}. The lack of late-type
giants close to the MBH had been reported before (see, e.g.,
discussion fo this topic in the review by \cite{Genzel:2010fk}), but
these more detailed studies were now considered to be incompatible
with the existence of a stellar cusp. This is the origin of the {\it
  missing cusp problem}.

The existence or not of a stellar cusp around the MBH at the GC has
implications for other galaxies as well, assuming that the Milky
Way's centre is representative for the nuclei of normal galaxies. Observational
confirmation or rejection of a stellar cusp would not only have
fundamental importance for our understanding of stellar
dynamics. It is also of particular importance for gravitational wave
astronomy. So called Extreme Mass Ratio Inspirals (EMRIs) are
considered the most exquisite probes of General Relativity and of the
related astrophysics of stellar remnant plus MBH systems
\cite{Amaro-Seoane:2007ve}. The possibility of observing EMRIs with
future gravitational wave observatories in space, such as LISA (Laser Interferometer Space Antenna )
\cite{Amaro-Seoane:2012ys,Amaro-Seoane:2013zr} or Taiji
\cite{Gong:2015zr} is directly linked to the question whether stellar
cusps exist around MBHs or not
\cite{Hopman:2005dn,Amaro-Seoane:2011qv}.

Motivated by the fundamental importance of this topic, we have
recently undertaken new observational and theoretical
studies. Observationally, we pushed the detection limit to fainter
stars. The observations were then compared to new, more realistic
Nbody simulations. We have found that a stellar cusp does indeed exist
at the GC and that its properties agree with our expectations. The
details of our work can be consulted in these three papers:
1. Gallego-Cano et al. (submitted to A\&A, arXiv:1701.03816);
2. Sch{\"o}del et al.  (submitted to A\&A, arXiv:1701.03817);
3. Baumgardt et al.  (submitted to A\&A, arXiv:1701.03818).  We
will briefly describe our findings in the following sections.

\section{Surface density of stars around the MBH Sgr\,A*}

\begin{figure}[htb]
\label{Fig:LF}
\includegraphics[width=\textwidth]{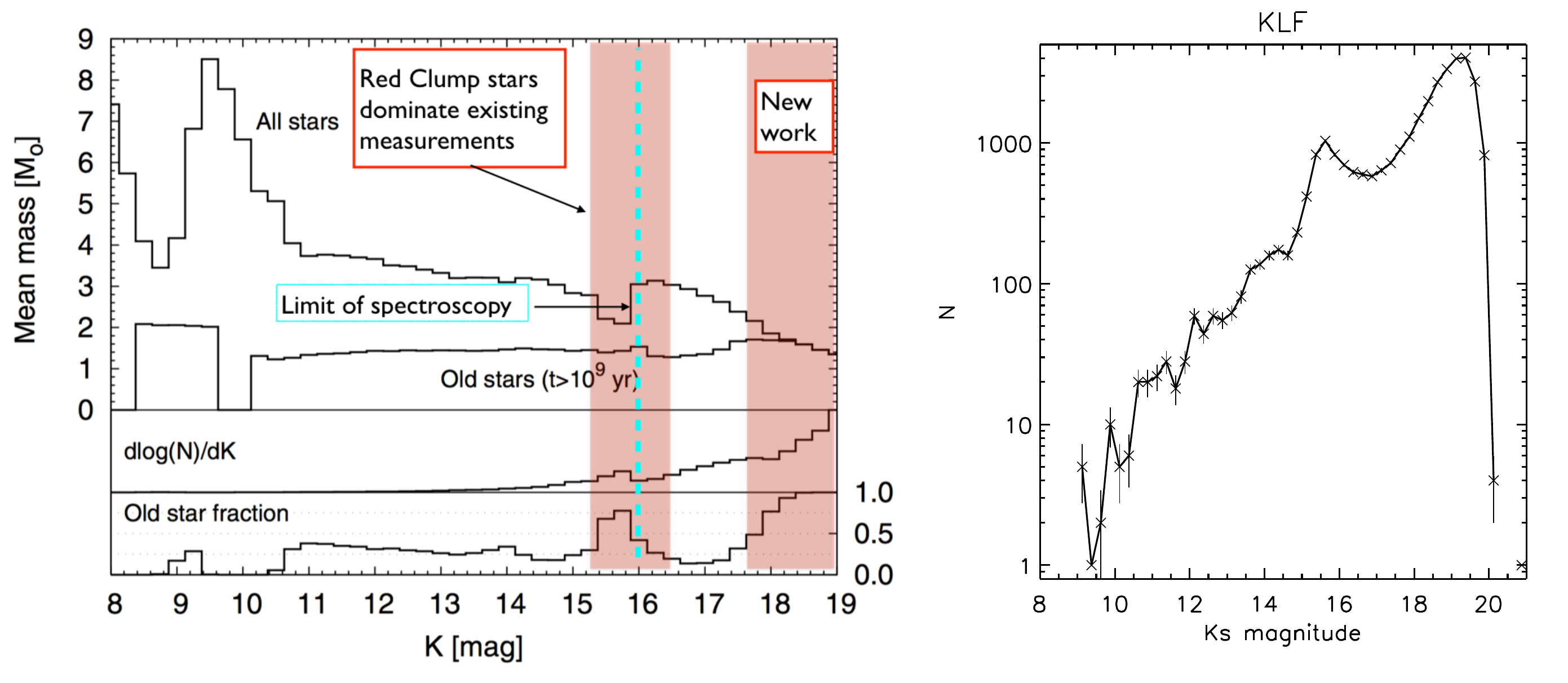}
\caption{Properties of the observable stellar population at the
  GC. {\it Left}: Simple population synthesis model of the GC stellar
  population, assuming continuous star formation, modified from
  Fig.\,16 in \cite{Schodel:2007tw} (Reproduced with permission from
  Astronomy \& Astrophysics, \copyright ESO). The distance modulus and
  extinction were added to the stellar magnitudes, which means that
  the magnitude values on the x-axis correspond to observed
  values. The upper panel shows the mean mass of all stars and of old
  stars (lifetime $> 1$\,Gyr) only. The middle panel shows a
  luminosity function. The lower panel displays the fraction of old
  stars in each magnitude interval. The completeness limit due to
  source crowding of previous studies was limited to
  $K_{s}\lesssim17.5$, while the sensitivity limit of spectroscopic
  studies is $K_{s}\approx16$. At $K_{s}<17.5$, number counts of old
  stars in the GC will therefore be dominated by horizontal branch/red
  clump (HB/RC) stars, core-helium burning giants. {\it Right}:
  $K_{s}$ luminosity function of stars within a projected distance of
  $R\lesssim10"$ of Sgr\,A*, extracted from a deep NACO/VLT S13 image.
  The bump centred at $K_{s}\approx15.75$ corresponds to the RC. The
  sharp drop in detected stars at $K_{s}>18$ is due to extreme source
  crowding, which impedes the detection of fainter stars. While RC
  stars and brighter giants have dominated the stellar density
  measurements of all previous work, we use measurements of resolved
  and unresolved stars at $Ks\gtrsim17.5$ in our new work.}
\end{figure}
 
Before interpreting the stellar surface and light densities, it is
important to have an approximate understanding of the stellar
population we expect to observe at the GC. While we are still far from
having a detailed understanding of the stellar population and its
formation history in the inner few parsecs of the Milky Way, the
following assumptions appear to be robust, according to our current
best knowledge: The NSC has undergone repeated star formation
episodes throughout its lifetime, with the most recent events
occurring a few $10^{6}$ and a few 10\,Myr ago. The majority of the
stars are old, with an estimated 80\% having formed more than 5\,Gyr
ago \cite{Genzel:2010fk,Pfuhl:2011uq}. 

The left panel in Fig.\,\ref{Fig:LF} shows a simple model of the
stellar population of the NSC
assuming a continuous star formation history (taken from Fig.\,16 in
\cite{Schodel:2007tw}). As mentioned above, it is very hard to
classify stars at the GC because photometric studies do usually not
cover a sufficient wavelength range and are usually not accurate and
precise enough to break the degeneracy between intrinsic stellar
colours and interstellar reddening and because spectroscopic studies
are extremely time-consuming and limited to the brightest
stars. Therefore, the diagram in the left panel in Fig.\,\ref{Fig:LF} is
valuable because it can give us an approximate idea of the ages and
masses of stars of a given observed brightness at the GC. If we are
interested in identifying a stellar cusp, then it is important to
focus on stars that are at least a few Gyr old. We can do this by selecting
spectroscopically classified stars and/or by limiting the brightness
range of the tracer population.  Previous studies of the NSC structure
(\cite{Buchholz:2009fk,Do:2009tg,Bartko:2010fk}) were dominated by
stars of K-band magnitudes brighter than $\sim$17.5, in particular by
Red Clump (RC) stars, helium core burning giants with observed K-band
magnitudes $15-16$ at the distance and extinction of the GC. 

RC stars are a convenient tracer population because they are, on
average, a few Gyr old. We can also see in the left panel in
Fig.\,\ref{Fig:LF} that stars of K-magnitudes 18 or fainter would also
be suitable tracer populations because they have low mean masses and
are therefore probably, on average, at least a few Gyr old.

In our new work, we focused explicitly on this faint, old, low mass
stellar population. We used K-band high angular resolution
observations from the camera NACO at the ESO VLT 8m telescope. Several
epochs of high quality images were stacked in order to increase the
sensitivity. Subsequently, correction factors for stellar crowding and
interstellar extinction were determined and applied to the
data. Systematic effects arising from different choices of key
parameters in the software packages that identify the stars in images
were explored. Finally, the stellar surface densities were determined
as a function of distance from the central MBH. Spherical symmetry of
the cluster was assumed, which is not strictly correct, but accurate
on a 10-20\% level. In parallel, we also explored the diffuse light
density from unresolved stars that could be measured in the images
after subtracting all identified, resolved stars and the emission from
diffuse, ionised gas. 

In order to constrain the structure of the entire NSC and to be able
to deproject the observed quantities, these data were complemented, at
projected distances $R\gtrsim1$\,pc from Sgr\,A*, by star
density/surface brightness measurements from other sources, with lower
angular resolution and/or sensitivity, but taken over a larger field
(\cite{Schodel:2014fk,Fritz:2016fj}). The latter were scaled to our
data in the overlapping regions. In Fig.\,\ref{Sigmanuker} we show the
surface brightness from unresolved stars from 0.01\,pc to 20\,pc as
well as the star counts for faint resolved stars (K-magnitudes
$\sim$18) and RC stars (K-magnitudes $\sim$15-16). The diffuse
flux from the unresolved stars traces sub-giants and main sequence
stars of K-band magnitude $19-20$, which have masses of about
$0.8-1.8$\,M$_{\odot}$. All these three tracer populations have
similar masses and can be old enough to be dynamically relaxed.

\section{Properties of the stellar cusp at the GC}

As Fig.\,\ref{Sigmanuker} shows, the three studied tracer populations
show a very similar distribution. The projected densities can all be
approximated well by single power-laws at projected distances $R<1$\,pc. Only
the brightest population, the red clump giants, show some systematic
deviations at around $R=0.2$\,pc and a possible decline at
$R<0.05$\,pc. The red line in Fig.\,\ref{Sigmanuker} indicates a fit
with a {\it Nuker} law \cite{Lauer:1995fk}:

\begin{figure}[htb]
\label{Sigmanuker}
\includegraphics[width=\textwidth]{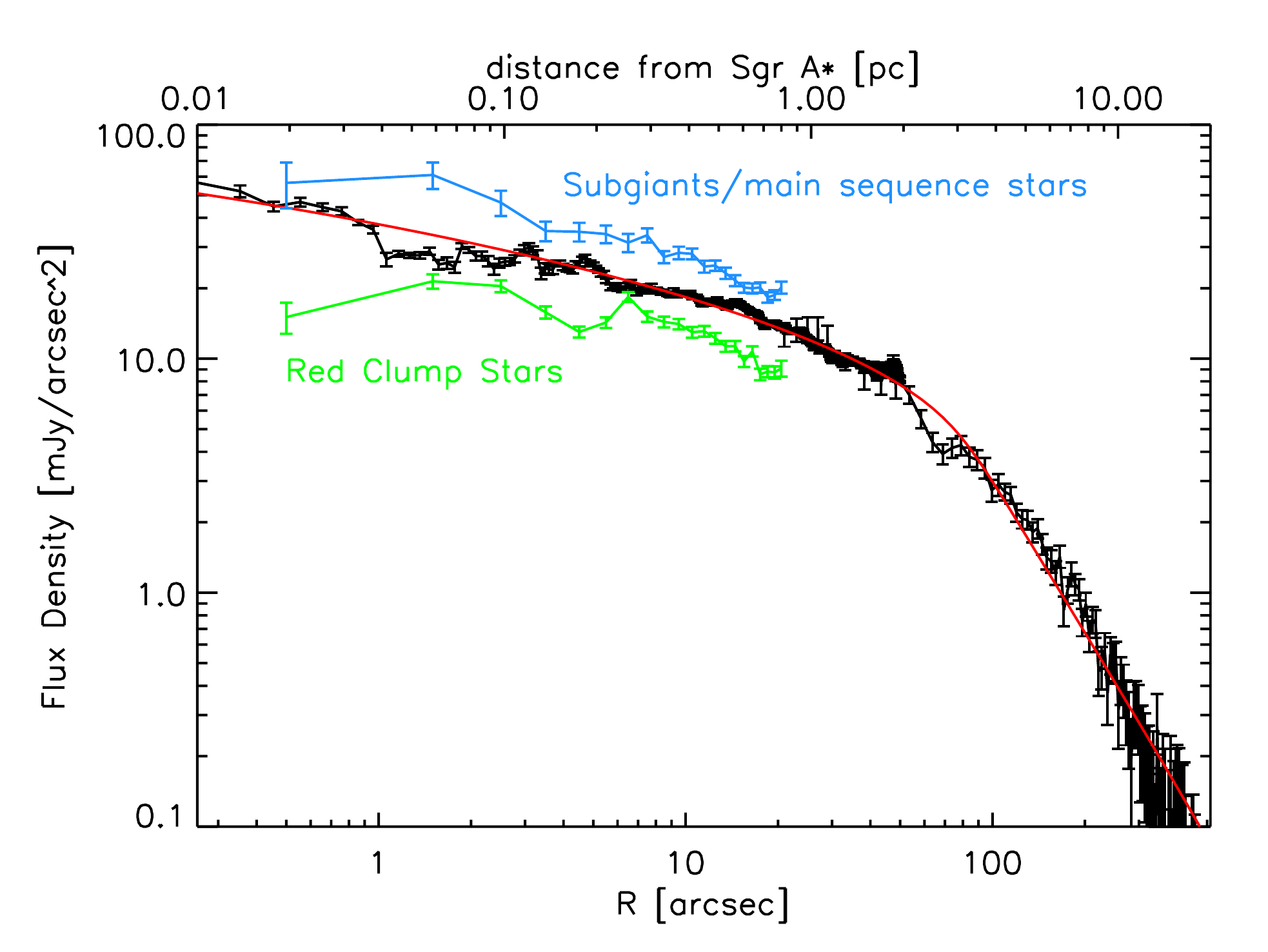}
\caption{Surface light/stellar number density at the GC plotted
  against the projected distance from Sgr\,A*. The black line is the
  surface light density from the unresolved stars (Sch{\"o}del et al.,
  submitted to A\&A,  arXiv:1701.03817). In order to constrain the shape of the NSC at
  distances $> 2$\,pc, we used the data from \cite{Fritz:2016fj} and
  scaled them in the overlap region to our data. A constant flux of
  6.4\,mJy\,arcsec$^{-2}$ was subtracted to remove the contamination
  by the surrounding nuclear bulge and Galactic bar, which can be
  approximated by a constant on these scales.  The red line is a
  best-fit Nuker model. The blue and green lines are scaled stellar
  number densities from Gallego-Cano et al. (submitted to A\&A, arXiv:1701.03816) for RC
  giants and faint subgiants/main sequence stars. }
\end{figure}
 
\begin{equation}
\rho(r) = \rho(r_{b})2^{(\beta-\gamma)/\alpha}\left(\frac{r}{r_{b}}\right)^{-\gamma}\left[1+\left(\frac{r}{r_{b}}\right)^{\alpha}\right]^{(\gamma-\beta)/\alpha},
\end{equation}

where $r$ is the 3D distance from the central MBH, $r_{b}$ is the
break radius, $\rho$ is the 3D density, $\gamma$ is the exponent of
the inner and $\beta$ the one of the outer power-law, and $\alpha$
defines the sharpness of the transition.  We carried out fits with
different parameters (e.g., for the value of $\alpha$ that was kept
fixed at a value of 10 during the fits) and data (data at large $R$ from
\cite{Schodel:2014fk} or \cite{Fritz:2016fj}) and determined mean
parameters and uncertainties from the resulting best-fit parameters of
the different tries (the formal uncertainties were always very
small). We obtain, for the unresolved stellar light
$r_{b,diffuse} = 3.2\pm0.2$\,pc, $\gamma_{diffuse}=1.16\pm0.02$,
$\beta_{diffuse}=3.2\pm0.3$, and
$\rho_{diffuse} (r_{b})=0.67\pm0.05$\,mJy\,arcsec$^{-3}$, for the star
counts of faint, resolved stars $r_{b,faint}=3.0 \pm 0.4$ pc
($77 \pm 10"$), $\gamma_{faint} = 1.29 \pm 0.02$,
$\beta_{faint} = 2.1 \pm 0.1$, and a density at the break radius of
$\rho_{faint} (r_{b})=3900 \pm 900$\,pc$^{-3}$, and for the RC stars
$r_{b,RC}=3.4 \pm 0.5$ pc ($110 \pm 13"$),
$\gamma_{RC} = 1.28 \pm 0.07$, $\beta_{RC} = 2.26 \pm 0.04$, and
$\rho_{RC} (r_{b})=1700 \pm 500$\,pc$^{-3}$.

We cannot directly compare the values of the normalisation parameters
$\rho$. As concerns $\beta$, it is not very well constrained. On the
other hand, the break radius $r_{b}$ is well constrained around 3\,pc
and the inner power-law index $\gamma$ as well. The break radius
corresponds roughly to the radius of influence of the Sgr\,A*, i.e.\ the
radius inside which we expect the stellar cusp to appear
\cite{Alexander:2005fk}. Both the stellar number and light densities
may suffer from not well constrained systematic biases (uncertainties
in sky background subtraction or completeness corrections etc.). If we
adopt conservative values from averaging the three values, we obtain
$r_{b}=3.2\pm0.2$ and $\gamma=1.24\pm0.07$, taking the standard
deviations as uncertainties. Obviously, the stellar density displays a
cusp and a core-like (flat) density law can be excluded with very high
confidence.

\section{Discussion and conclusions}

We find that the stars at the GC follow a cusp-like density
distribution inside the radius of influence of the central MBH. Our
findings resolve the ambiguity from previous work. The cusp is less
steep than what is expected for a single age population that is older
than the relaxation time. In the latter case one would expect to
observe a so-called Bahcall-Wolf cusp with $\gamma\geq1.5$. Numerical
experiments were performed by Baumgardt et
al. (arXiv:1701.03818) to interpret our data. They performed Nbody simulations of a more
realistic NSC, assuming repeated star formation episodes every
1\,Gyr. This means that most stars will not have had the time to fully
relax dynamically, which will flatten the cusp. The results of the
simulations agree very well with the data. 

There appears to be a lack of giant stars at small projected distances
of $R\lesssim0.1$\,pc from Sagittarius\,A*. The missing few dozens of
giants could be explained if their envelopes were removed in the past,
thus rendering them invisible to observations. While collisions between
stars or between stellar remnants and stars are probably not effective
enough, collisions with high-density clumps in a formerly existing
star-forming gas disc around Sagittarius\,A* provide a viable
mechanism \cite{Amaro-Seoane:2014fk}. We know that such a disc must
have existed in the past because of the presence of many young,
massive stars in this region that move in a coherent, disc-like
pattern (e.g., (e.g.,
\cite{Beloborodov:2006ff,Lu:2009bl,Genzel:2010fk}).

We conclude that a stellar cusp exists around Sagittarius\,A* and that
its properties agree with what we expect theoretically. Densities in
excess of a few $10^{7}$\,M$_{\odot}$\,pc$^{-3}$ are reached at
distances $r<0.01$\,pc from the central black hole (Sch\"odel et al.,
submitted to A\&A, arXiv:1701.03817).  This has significant
implications for gravitational wave astronomy. The existence of a
stellar cusp in the Milky Way implies the existence of such structures
in other galaxies with similar or smaller MBHs. Therefore we can
expect to observe significant numbers of EMRIs with future space-based
gravitational wave observatories
\cite{Hopman:2005dn,Amaro-Seoane:2012ty,Amaro-Seoane:2012ys,Amaro-Seoane:2013zr}.

\ack 

The research leading to these results has received funding from the
European Research Council under the European Union's Seventh Framework
Programme (FP7/2007-2013) / ERC grant agreement n$^{\circ}$
[614922]. PAS acknowledges support from the Ram{\'o}n y Cajal
Programme of the Spanish Ministerio de Economía, Industria y
Competitividad.

\bibliography{/Users/rainer/Documents/BibDesk/BibGC}

\end{document}